\documentclass[aps,showpacs,twocolumn]{revtex4}
\usepackage{graphicx}
\usepackage{dcolumn}
\usepackage{bm}
\usepackage{latexsym}
\begin{document}
\newcommand{\A}{{\mathcal{A}}}
\newcommand{\dA}{\delta{\mathcal{A}}}
\newcommand{\oR}{\Omega_r h^2}
\newcommand{\og}{\Omega_{\gamma} h^2}
\newcommand{\om}{\Omega_m h^2}
\newcommand{\ok}{\Omega_k h^2}
\newcommand{\ob}{\Omega_b h^2}
\newcommand{\ode}{\Omega_{de} h^2}
\newcommand{\h}{\hat{H}}

\title{Cosmic vector for dark energy: constraints from SN, CMB and BAO}
\author{Jose Beltr\'an Jim\'enez$^a$, Ruth Lazkoz$^b$ and Antonio L. Maroto$^a$}
\affiliation{$^a$Departamento de  F\'{\i}sica Te\'orica,
 Universidad Complutense de
  Madrid, 28040 Madrid, Spain.\\
$^b$Fisika Teorikoa, Zientzia eta Teknologiaren Fakultatea, Euskal
Herriko Univertsitatea, 644 Posta Kutxatila, 48080 Bilbao, Spain.}

\date{\today}

\begin{abstract}
It has been recently shown that the presence of a vector field
over cosmological scales
could explain the observed accelerated expansion
of the universe without introducing neither new scales nor
unnatural initial conditions in the early universe,
thus avoiding the coincidence
problem.
Here, we present a detailed analysis of the constraints imposed
by  SNIa, CMB and BAO data on the vector dark energy model
with general spatial curvature. We find  that contrary to standard
cosmology, CMB data excludes a flat universe for this model and, in fact,
predicts a closed geometry for the spatial sections.
We see that CMB and SNIa Gold data are perfectly compatible at the 1-sigma
level, however SNIa Union dataset exhibits a 3-sigma tension with
CMB. The same level of tension is also found  between SNIa and
BAO measurements.

\end{abstract}


\maketitle

\section{Introduction}

More than ten years have passed since the first indications of the
accelerated expansion of the universe  \cite{SN1998} and still
today it remains as one of the most intriguing problems in
cosmology. Moreover, the accelerated expansion has been confirmed
during the last decade by many different probes, mainly through
measurements of Type Ia supernovae,  CMB temperature power
spectrum and the Baryon Acoustic Oscillations (BAO)
\cite{gold,Union,CMB}. Since Einstein's gravity predicts that a
universe containing (baryonic and dark) matter and radiation
should be decelerating rather than accelerating, these
observations could be signalling either the breakdown of General
Relativity at cosmological scales or the existence of some sort of
non-ordinary energy with negative pressure known as dark energy.

The simplest model describing dark energy is the existence of a
cosmological constant. Although it fits  observations with very
good precision, it suffers from theoretical problems since the
inferred scale for the cosmological constant turns out to be
around $10^{-3}$eV, which is a very tiny value compared to the
gravitational scale set by Newton's constant $G \simeq M_P^{-2}$
with $M_P\sim 10^{19}$GeV. This poses a problem of naturalness
because of the existence of two scales in the theory which differ
in many orders of magnitude. Moreover, a related problem arises
because the amount of energy density stored in the form of dark energy
is comparable to that stored in the form of matter at the present
epoch, in spite of having evolved very differently in the past.
Thus, in order to get these problems around several models have
been proposed which are mostly based on either cosmological scalar
fields \cite{quintessence,k-essence} or infrared modifications of
the gravitational action \cite{f(R),DGP} (see \cite{Copeland} for
a review on dark energy models). However, none of these models
succeeded in solving the previously addressed problems of the
cosmological constant because they either introduce new
dimensional scales in the action or unnatural initial conditions
to get the right acceleration at the right time.

In a previous paper \cite{cosmicvector} we showed that a model
based on the dynamics of a vector field on cosmological scales can
give rise to a period of accelerated expansion without introducing
neither new dimensional scales nor unnatural initial conditions,
thus avoiding  fine-tuning or coincidence problems. Although the
existence of periods of accelerated expansion for vector field
models was already known \cite{Kiselev, Boehmer}, the
model proposed in \cite{cosmicvector} consists just of the
simplest kinetic terms for the vector field containing two fields
and two derivatives and without any potential for the field. In
fact, in this paper we shall show that such a model is nothing but
a gauge field with a gauge fixing term coupled to the Ricci
tensor. Moreover, very recently \cite{EM} it has been also shown
that the own electromagnetic field could be a natural
candidate for dark energy.

In order to constrain dark energy models from observations, we
typically use distance indicators so that we can confront distance
measurements  to the corresponding  model predictions
\cite{DEcomparisons}. To do that we can resort to two different
types of objects, namely: standard candles and standard rulers.
Standard candles are objects of known intrinsic luminosity, so
that  the corresponding comoving distance can be determined. That
way, it is possible to reconstruct the Hubble expansion rate
 by searching this sort of objects at
different redshifts. The most important class of such indicators are
Type Ia supernovae. On the other
hand, standard rulers are objects whose comoving size is known, so
that we can measure the angular distance and, therefore, compare
to that predicted by the dark energy model. A  well-known example is
 the sound horizon size at  the last scattering surface. This
standard ruler can be measured directly from the CMB temperature
power spectrum and, also, from Baryon Acoustic Oscillations (BAO)
through the matter power spectrum at low redshift.

The paper is organized as follows: in section I we introduce the
vector model for dark energy and derive all the necessary
equations for the rest of the paper. In section II we show how the
different distance indicators will be used in order to obtain
constraints for the model. Finally, section III is devoted to the
results obtained from the analysis.

\section{Vector dark energy}
The action proposed in \cite{cosmicvector} to describe dark energy
is the following:
\begin{eqnarray}
S&=&\int d^4x \sqrt{-g}\left(-\frac{R}{16\pi G}\right.\nonumber\\
&-&\left.\frac{1}{2}\nabla_\mu A_\nu\nabla^\mu A^\nu +\frac{1}{2}
R_{\mu\nu}A^\mu A^\nu\right) \label{action}
\end{eqnarray}
However, this action can be written in a more suggestive form by
integrating by parts and taking into account that $R_{\mu\nu}A^\mu
A^\nu=(\nabla_\mu A^\mu)^2-\nabla_\mu A_\nu\nabla^\nu A^\mu$.
Then, the action for the vector field can be expressed as:
\begin{eqnarray}
S&=&\int d^4x \sqrt{-g}\left[-\frac{R}{16\pi G}\right.\nonumber\\
&-&\left.\frac{1}{4}F_{\mu\nu}F^{\mu\nu}-\frac{1}{2}\left(\nabla_\mu
A^\mu\right)^2+ R_{\mu\nu}A^\mu A^\nu\right] \label{action}
\end{eqnarray}
In this form, it becomes apparent that the theory is that of a
gauge vector field with a gauge-fixing term (in the Feynman gauge)
and coupled to the Ricci tensor. This coupling provides an
effective mass term driven by gravity.

Now, we take variations with respect to the metric and the vector
field to obtain the Einstein's and vector field equations
respectively:
\begin{eqnarray}
R_{\mu\nu}-\frac{1}{2}R g_{\mu\nu}&=&8\pi G
(T_{\mu\nu}+T_{\mu\nu}^A)\label{eqE}\\
\nabla_\nu\nabla^\nu A_\mu + R_{\mu\nu}A^\nu&=&0 \label{eqA}
\end{eqnarray}
where $T_{\mu\nu}$ is the conserved energy-momentum tensor for
matter and radiation (and all other possible components) and
$T_{\mu\nu}^A$ is the energy-momentum tensor for the vector field.
We consider the case in which the vector field is homogeneous and
only has time-component, i.e., $A_\mu=(A_0(t),0,0,0)$.
However, unlike in the previous paper \cite{cosmicvector}, here we
will consider the effects of the curvature so the metric is given
by:
\begin{equation}
ds^2=dt^2-a(t)^2\left[\frac{dr^2}{1-kr^2}+r^2\left(d\theta^2+\sin^2\theta
d\phi^2\right)\right]\label{metric}
\end{equation}
In this metric, equation (\ref{eqA}) for the homogeneous time-component of the vector field reads:
\begin{eqnarray}
\ddot{A}_0+3H\dot{A}_0-3\left(2H^2+\dot{H}\right)A_0=0
\end{eqnarray}
where $H=\dot a/a$ is the Hubble parameter. For our purposes on
this work it will be useful to express the latter equation in
terms of the redshift $z=\frac{1}{a}-1$ as follows:
\begin{equation}
\frac{d^2A_0}{dz^2}+\frac{1}{(1+z)H(z)}\frac{d}{dz}\left[\frac{H(z)}{(1+z)^2}
\right]
\frac{d}{dz}\left[(1+z)^3A_0\right]=0\label{fieldeq0}
\end{equation}
This equation can be easily solved when the universe is dominated
by radiation or matter, assuming that the contribution of the vector field is negligible. In those epochs the Hubble parameter is given by $H=p/t=p(1+z)^{1/p}$ with $p=1/2$ for radiation and $p=2/3$ for
matter and, according to (\ref{fieldeq0}), the vector field evolves
as:
\begin{equation}
A_0(z)=A_0^+(1+z)^{\alpha_+}+A_0^-(1+z)^{\alpha_-}\label{fieldsol}
\end{equation}
with $A_0^\pm$ constants of integration and $\alpha_{\pm}=(1\pm
1)/2$ in the radiation era, and $\alpha_{\pm}=(3\pm\sqrt{33})/4$
in the matter era. Notice that, since the vector field is constant
during the radiation era, we can set the initial conditions at any
time during that epoch without modifying the evolution of the
universe.

\begin{figure*}[htp]
\begin{center}
\includegraphics[width=17cm]{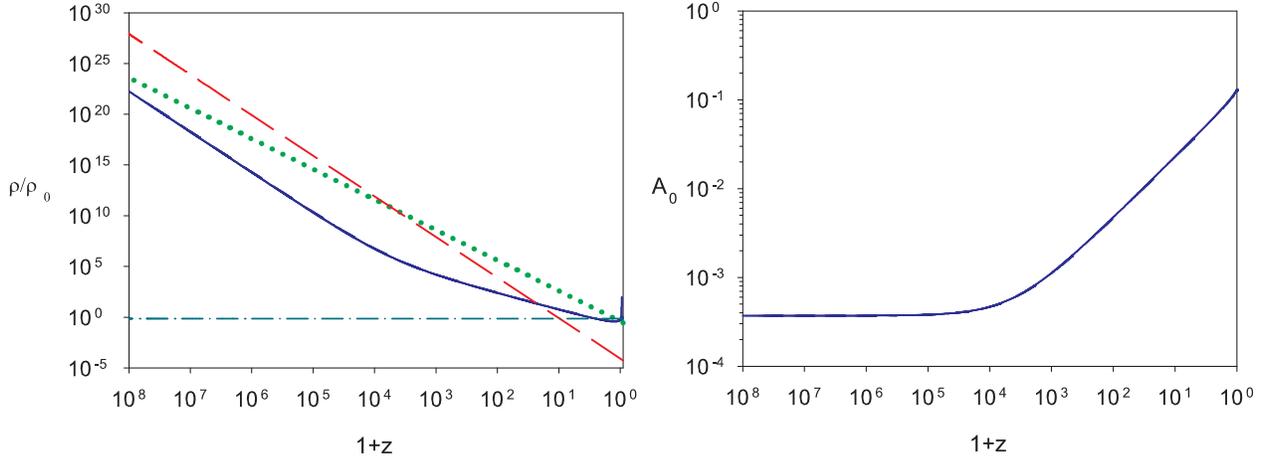}\caption{\small Left panel:
evolution of
energy densities. Dashed (red) for radiation, dotted (green) for
matter and solid (blue) for vector dark energy. We also show for
comparison the cosmological constant energy density in
dashed-dotted line. We see the rapid growth of dark energy
contribution at late times approaching the final singularity.
Right panel: cosmological evolution of the vector field. In these
two plots we see the unimportance of the time at which we set the
initial conditions due to the fact that both the vector field and
the fraction of dark energy density are constant in the early
universe.}\label{cosmevol}
\end{center}
\end {figure*}

On one hand, the $(^0\; _0)$ component of Einstein's equations can
be written as:
\begin{equation}
\frac{H^2}{H_0^2}=\Omega_m(1+z)^3+\Omega_r(1+z)^4+\Omega_k(1+z)^2+\rho_A(z)
\label{Friedmann}
\end{equation}
where $H_0$ is the present value of the Hubble parameter, which is
usually expressed as $H_0=100 \,h \,$km s$^{-1}$Mpc$^{-1}$, $\Omega_m$
and $\Omega_r$ are the density parameters corresponding to
matter and radiation respectively, $\Omega_k=-\frac{k}{H_0^2}$ and
$\rho_A$ is the energy density of the vector field, whose
expression is:
\begin{equation}
\rho_{A}=\frac{H^2}{H_0^2}\left[\frac{1}{2}A_0^2-(1+z)A_0\frac{dA_0}{dz}
-\frac{1}{6}(1+z)^2\left(\frac{dA_0}{dz}\right)^2\right]
\end{equation}
Moreover, we measure the vector field in units of the reduced
Planck mass $\tilde M_P=1/\sqrt{8\pi G}$. We can obtain the evolution of
this energy density in the radiation and matter dominated eras by
using the growing solution in (\ref{fieldsol}) for the vector
field to give:
\begin{equation}
\rho_A=\rho_{A0}(1+z)^\kappa\label{rhoevo}
\end{equation}
where $\kappa=4$ in the radiation era and
$\kappa=(9-\sqrt{33})/2\simeq1.63$ in the matter era. Notice that
the energy density scales as radiation in the early universe so
that $\rho_A/\rho_R$ is constant in that epoch (see in Fig.
\ref{cosmevol}). Therefore, we obtain once again that the time at
which we set the initial conditions in the early universe for the
vector field lacks importance, since its fraction of energy
density is constant during that epoch.

Finally, when the vector field becomes dominant the Universe
suffers a Type III singularity \cite{Nojiri} in which its evolution ends at a
finite time $t_{end}$ but with a finite size $a_{end}$. Moreover,
as we approach the final singularity we have $\rho_{DE}\rightarrow
\infty$ and $p_{DE}\rightarrow -\infty$, with $w_{DE}\rightarrow
-\infty$, whereas the vector field takes a finite value.

On the other hand, the $(^i\;_i)$ component of Einstein's
equations is:
\begin{equation}
\frac{H^2}{H_0^2}\left[3-\frac{1+z}{H}\frac{dH}{dz}\right]=
\Omega_r(1+z)^4+\Omega_k(1+z)^2-p_A(z)\label{ecacc}
\end{equation}
where we have used that $p_r=\frac{1}{3}\rho_r$ and $p_m=0$ and
the pressure of the vector field is given by:
\begin{eqnarray}
p_A(z)&=&-\frac{H^2}{H_0^2}\left\{3\left[\frac{5}{2}-\frac{4}{3}(1+z)\frac{1}{H}
\frac{dH}{dz}\right]A_0^2\right.\nonumber\\
&+&(1+z)A_0\frac{dA_0}{dz}
+\left.\frac{3}{2}(1+z)^2\left(\frac{dA_0}{dz}\right)^2\right\}
\end{eqnarray}

In order to perform the analysis in next sections it will be
convenient to write  equations (\ref{Friedmann}) and (\ref{ecacc})
in terms of $\{\oR,\om,\ok\}$ as follows:
\begin{equation}
\h^2=\om(1+z)^3+\oR(1+z)^4 +\ok(1+z)^2+\rho_A(z)
\label{Friedmann2}
\end{equation}
\begin{equation}
\h\left[3-\frac{1+z}{\h}\frac{d\h}{dz}\right]=\oR(1+z)^4
+\ok(1+z)^2-p_A(z)\label{acceq2}
\end{equation}
where $\h\equiv H/(100 \,$km s$^{-1}$\,Mpc$^{-1})$, i.e., $\h(z=0)=h$. Note
that neither $\rho_A$ nor $p_A$ depend on the normalization of the
Hubble parameter. Moreover, $\oR$ contains the contribution of
photons as long as relativistic neutrinos,
\begin{equation}
\oR=\og(1+0.2271N_{eff})
\end{equation}
with $N_{eff}=3.04$ the effective number of neutrino species and
$\og=2.469\times10^{-5}$ for $T_{CMB}=2.725$.

The model is completely determined once we fix the set of
parameters $\{\Omega_m,\Omega_k,A_{rad}\}$ so that the model has three free parameters. To confront the model to SN and BAO dataset we only need to integrate the system of equations up to redshift $\sim 2$ whereas CMB dataset requires to obtain the solution up to the last scattering surface so that the method to solve the equations will be different for each case. The present value of
the Hubble expansion rate is no longer a free parameter in this model because it can be obtained in terms of the previous parameters after integrating the equaions. In fact, we could take $\{\Omega_m,\Omega_k,h\}$ as independent parameters and,
therefore, $A_{rad}$ would already be determined, although this approach is more difficult to implement numerically. Notice that this model contains exactly the same number of parameters as $\Lambda$CDM.

\section{Likelihood calculations}
In this section we shall explain the procedure followed to
confront the vector dark energy model to the different distance
indicators.

\subsection{SN}
The apparent magnitude of a supernova placed at a
given redshift $z$ is related to the expansion history of the
universe through the distance modulus:
\begin{equation}
\mu\equiv m-M=5\log D_L -5\log h+\mu_0
\end{equation}
where $m$ and $M$ are the apparent and absolute magnitudes
respectively, $\mu_0=42.38$ and $D_L=H_0d_L$ with $d_L$ the
luminosity distance $d_L=(1+z)r(z)$, being $r(z)$ the comoving
distance, given for the metric (\ref{metric}) by:
\begin{equation}
r(z)=\frac{1}{H_0\sqrt{|\Omega_k|}}\;S_k\left[\sqrt{|\Omega_k|}\int_0^z\frac{H_0
}{H(z')}dz'\right]
\end{equation}
with $S_k[x]=\sin x,x,\sinh x$ for
$\Omega_k<0,\Omega_k=0,\Omega_k>0$ respectively.

Then, to confront the model to each supernovae data set we
construct the corresponding $\chi^2$ estimator:
\begin{equation}
\chi^2_{SN}=\sum_{i=1}^N\frac{\left(\mu(z_i;\Omega_m,\Omega_k,h)-\mu_i\right)^2}
{\sigma_i^2}\label{chiSNh}
\end{equation}
which must be marginalized over $h$ in order to obtain the constraints
on the parameters $\Omega_m$ and $\Omega_k$.

In order to calculate $\chi^2_{SN}$, we use the fact that the SNIa
dataset corresponds to redshifts below 2 so that we
can neglect the contribution from radiation in Einstein's
equations. With this in mind, we solve numerically the system of equations (\ref{fieldeq0}) and (\ref{ecacc}) for $H/H_0$ and
$A_0$. As this system is of second order with respect to $A_0$ and first order with respect to $H/H_0$ we need to set the intial values of $A_0$, $dA_0/dz$ and $H/H_0$. However, these three initial values are related by means of Friedmann equation
(\ref{Friedmann}) so that we can obtain the initial value for $H/H_0$ in terms of the initial values of $A_0$ and its derivative. On the other hand, as we know the analytic solution of the vector field
in the matter dominated era as that given in (\ref{fieldsol}) , we can relate the initial value of the
derivative of the vector field to the initial value of
the vector field (neglecting the decaying mode). Therefore, we only need to give the initial value for $A_0$ in order to set the initial conditions and we are left with $A_0^{ini}$, $\Omega_m$ and $\Omega_k$ as free parameters in terms of which we obtain the corresponding
$\chi^2_{SN}$ estimator. Therefore, we shall use:
\begin{equation}
\chi^2_{SN}=\sum_{i=1}^N\frac{\left(\mu(z_i;\Omega_m,\Omega_k,A_{ini})-\mu_i\right)^2}
{\sigma_i^2}
\end{equation}
instead of (\ref{chiSNh}) and marginalize over $A_{ini}$.

In this work we have used two sets of supernovae: the Gold set
\cite{gold} and the more recent Union set \cite{Union}.

\subsection{BAO}

BAO measurements provide the following distance ratios
\cite{Percival}:
\begin{eqnarray}
{\bf V}_{BAO}\equiv\left(\begin{array}{c}
\frac{r_s(z_d)}{D_V(0.2)}\\\frac{r_s(z_d)}{D_V(0.35)}\end{array}\right)=
\left(\begin{array}{c}0.1980\pm0.0058\\0.1094\pm
0.0033\end{array}\right),\label{BAO2}
\end{eqnarray}
where $r_s(z)$ is the sound horizon size given by:
\begin{equation}
r_s(z)=\frac{1}{\sqrt{3}}\int_0^{\frac{1}{1+z}}
\frac{da}{a^2H(a)\sqrt{\left(1+\frac{3\ob}{4\og}a\right)}}
\end{equation}
and
\begin{equation}
D_V(z)=\left[r^2(z)\frac{z}{H}\right]^{1/3}
\end{equation}
is the dilation scale. Finally, $z_d$ is the drag epoch at which
baryons were released from photons and which can be calculated by
using the fitting formula \cite{Huzdrag}:
\begin{equation}
z_d=\frac{1291(\om)^{0.251}}{1+0.659(\om)^{0.828}}\left[1+b_1(\ob)^{b_2}\right]
\end{equation}
with
\begin{eqnarray}
b_1&=&0.313(\om)^{-0.419}\left[1+0.607(\om)^{0.674}\right]\\
b_2&=&0.238(\om)^{0.223}.
\end{eqnarray}
Then, we define the BAO array:
\begin{eqnarray}
{\bf X}_{BAO}\equiv\left(\begin{array}{c}
\frac{r_s(z_d)}{D_V(0.2)}-1.980\\\frac{r_s(z_d)}{D_V(0.35)}-0.1094
\end{array}\right),
\end{eqnarray}
so that:
\begin{equation}
\chi_{BAO}^2={\bf X}_{BAO}^T{\bf C}_{BAO}^{-1}{\bf
X_{BAO}}.\label{chibao}
\end{equation}
In this expression, the inverse covariance matrix is
\begin{eqnarray}
{\bf C}_{BAO}^{-1}=\left(
\begin{array}{ccc}
35059&-24031\\
-24031&108300
\end{array}\right).
\end{eqnarray}
The procedure we follow in this case is analogous to that used for the
SNIa analysis, although, as $\chi^2_{BAO}$ depends on the amount of baryons
$\Omega_b$, we also need to marginalize over this parameter.

\subsection{CMB}
Following \cite{Komatsu}, we use the distance priors method to
confront dark energy models to CMB data \cite{Wang}. This method
uses two distance ratios measured by means of the CMB temperature
power spectrum:
\begin{itemize}
\item The "acoustic scale", which measures the ratio of the
angular diameter distance to the decoupling epoch and the comoving
sound horizon size at decoupling epoch. This first distance ratio
can be expressed as:
\begin{equation}
l_A\equiv\frac{\pi r(z_*)}{r_s(z_*)}
\end{equation}
Moreover, we use the fitting formula of $z_*$ proposed in
\cite{Huzstar}:
\begin{eqnarray}
z_*=&&1048\left[1+0.00124(\ob)^{-0.738}\right]\nonumber\\
&\times&\left[1+g_1(\om)^{g_2}\right]
\end{eqnarray}
with
\begin{eqnarray}
g_1=\frac{0.0783(\ob)^{-0.238}}{1+39.5(\ob)^{0.763}}\\
g_2=\frac{0.560}{1+21.1(\ob)^{1.81}}
\end{eqnarray}
\item The second distance ratio measures the ratio of the angular
diameter distance and the Hubble radio at the decoupling time. It
is usually called the "shift parameter" and can be expressed as
\begin{equation}
R=\sqrt{\Omega_m H_0^2}r(z_*)
\end{equation}
\end{itemize}

\begin{figure*}[htp]
\begin{center}
\includegraphics[width=17cm]{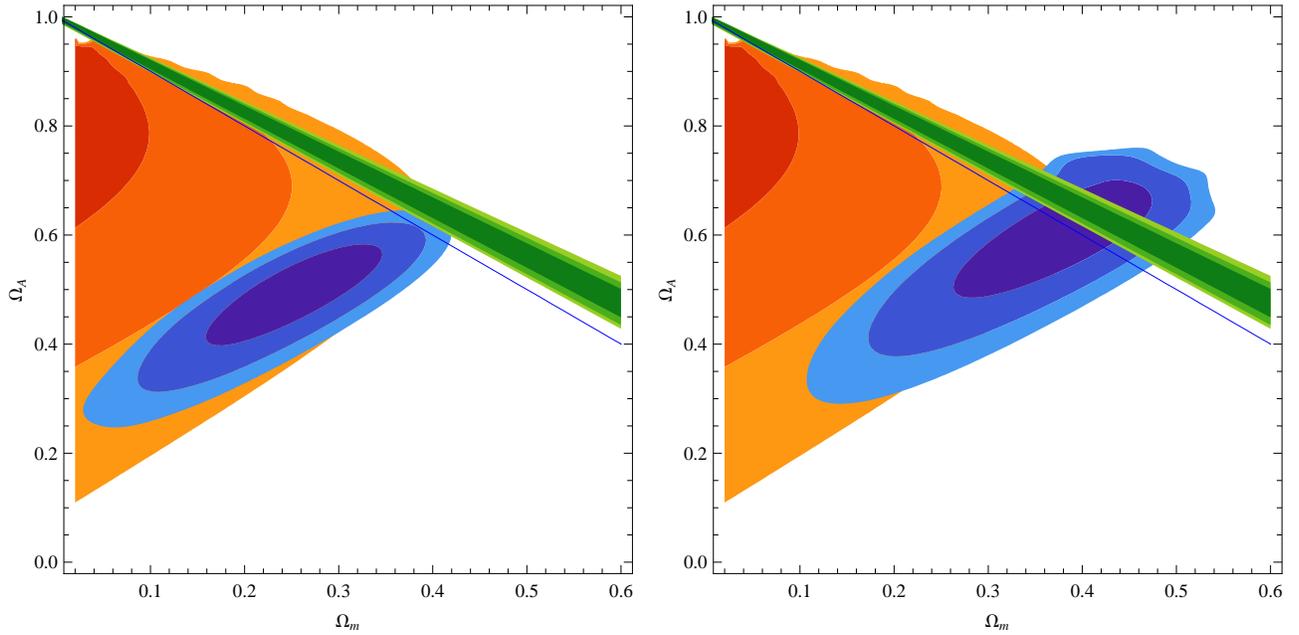} \caption{\small  In this
plots we show the $68\%$, $95\%$ and $99\%$ C.L. regions for BAO
(orange), CMB (green) and SNIa (blue). We show the contours
obtained for both the Union data set (left) and the Gold data set
(right). The blue line corresponds to a flat
universe.}\label{contours}\end{center}
\end{figure*}

\begin{figure}[htp]
\begin{center}
\includegraphics[width=8cm]{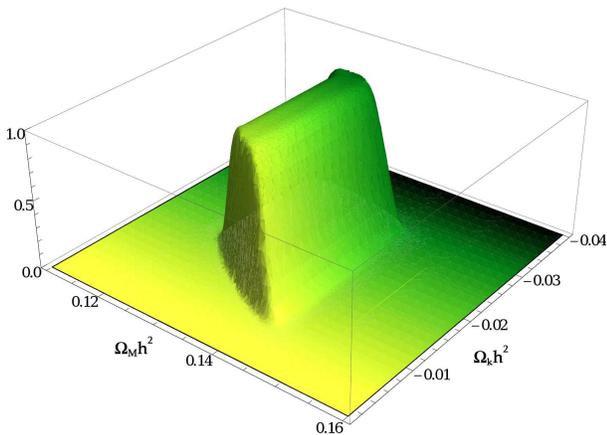} \caption{\small
In this plot we show the likelihood obtained from the CMB dataset for
the vector field model. We can see that a flat universe is clearly
ruled out and a closed geometry for the spatial sections is
strongly favored. }\label{likelihood}\end{center}
\end{figure}

The values reported in \cite{Komatsu} for these distance priors
are:
\begin{eqnarray}
{\bf V}_{CMB}\equiv\left(\begin{array}{c}
l_A(z_*)\\R(z_*)\\z_*\end{array}\right)= \left(\begin{array}{c}
302.10\pm 0.86\\1.710\pm0.019\\1090.04\pm0.93\end{array}\right)
\end{eqnarray}
with the following inverse of the covariance matrix
\begin{eqnarray}
{\bf C}_{CMB}^{-1}=\left(
\begin{array}{ccc}
1.800 &27.968& -1.103\\
27.968 &5667.577& -92.263\\
-1.103 &-92.263 &2.923
\end{array}\right).
\end{eqnarray}
Then, we define the CMB array as
\begin{eqnarray}
{\bf X}_{CMB}=\left(\begin{array}{c}
l_A-302.10\\R-1.710\\z_*-1090.04\end{array}\right),
\end{eqnarray}
so that:
\begin{equation}
\chi_{CMB}^2={\bf X}_{CMB}^T{\bf C}_{CMB}^{-1}{\bf X_{CMB}}.
\end{equation}

The procedure we follow in this case is somewhat different to
that used in the previous sections. The main difference comes from the
fact that CMB distance priors are
evaluated at a time when radiation is important so that we cannot neglect
its contribution in Einstein's equations anymore. To simplify numerical
calculations we use equations (\ref{fieldeq0}), (\ref{Friedmann2}) and
(\ref{acceq2}). Notice that, unlike  the SN and BAO approach, from these
equations we obtain the Hubble expansion rate normalized to
$100$ km s$^{-1}$Mpc$^{-1}$ so that $\h(z=0)=h$. Thus, for given
$\{\om,\ok\}$ we use (\ref{Friedmann2}) to relate the initial
condition for the Hubble expansion rate to the intial condition of
the vector field (that we shall name $A_{rad}$) and, then, solve
numerically (\ref{fieldeq0}) and
(\ref{acceq2}). Since the initial conditions are set in the radiation-dominated
era when, according to (\ref{fieldsol}), the vector field is constant, the
initial condition for the derivative of the vector field is set to zero.
Moreover, the constancy of the vector field during that epoch eliminates
the dependency on the time at which we place the initial conditions,
i.e., $A_{rad}$ does not depend on $z_{ini}$. That way, we obtain the
expansion rate $\h(z)$
that will allow us to compute the distance indicators described
above in terms of $\{\om,\ok, A_{rad}\}$ (notice that such
indicators do not depend on the normalization of the Hubble expansion
rate). Hence, we can compute the corresponding $\chi^2_{CMB}$
which will depend on $\{\om, \ok, \ob, A_{rad}\}$ and, following the
prescription given in \cite{Komatsu}, we marginalize over $\ob$
and $A_{rad}$ (which is equivalent to marginalize over $h$) and
use the resulting marginalized likelihood to obtain the
corresponding contours.

Since CMB distance priors were derived in \cite{Komatsu} assuming that
dark energy was not important at decoupling time ($z_*\simeq 1090$)
and given that the vector field model does not produce a significant amount of
dark energy
at high redshifts, these priors are, in principle, applicable in this case.

\section{Results}

In this section we present the results obtained after confronting
the model given by the action (\ref{action}) with the tests
explained above. We have also performed the analysis for a
$\Lambda$CDM model for comparison.

Using the Gold data set we obtain a best fit for $\Omega_m=0.385$
and $\Omega_A=0.611$ with $\chi_{min}^2=172.92$, which is the same
value found in \cite{cosmicvector} where we imposed  flat spatial
sections. This is understandable because, from the above values of
$\Omega_m$ and $\Omega_A$ we obtain $\Omega_k=0.0043$ so that the
best fit is very close to the flat case. However the 1$\sigma$
contour allows both open and close universes and, unlike
\cite{cosmicvector}, a wide range of values for $\Omega_m$ and
$\Omega_A$ is within the 1$\sigma$ region, as we can see in Fig.
(\ref{contours}). For a $\Lambda$CDM model with non-vanishing
curvature we obtain the best fit for $\Omega_m=0.46$ and
$\Omega_\Lambda=0.98$ with $\chi_{min}^2=175.04$ so we still
obtain a better fit to the Gold data set than $\Lambda$CDM. On the
other hand, the best fit obtained for the vector dark energy model
from the Union data set corresponds to $\Omega_m=0.260$ and
$\Omega_A=0.503$ with $\chi_{min}^2=311.96$. From Fig.
(\ref{contours}) we see that this data set favors an open universe
for this model, being the flat case at more than 2$\sigma$. For
$\Lambda$CDM the best fit happens for $\Omega_m=0.41$ and
$\Omega_\Lambda=0.93$, being $\chi_{min}^2=310.23$, which is lower
than that obtained for the vector field. This effect is probably
due to the SNLS points contained in the Union data set which, as
it is shown in \cite{cosmicvector}, favor $\Lambda$CDM over the
vector field model at more than 2$\sigma$ in the flat case.
However, when the flatness assumption is dropped, $\Lambda$CDM
fits the Union data set better than the vector field model only at
less than 1$\sigma$.

Concerning BAO dataset, it favors an open universe with a small
amount of matter for the vector field model, as we see in Fig.
(\ref{contours}). Moreover, the compatibility of these data with
SNIa data is only at the 3$\sigma$ level. However, it is worth
mentioning that these distance indicators are obtained after
analyzing the actual observational data with $\Lambda$CDM as
fiducial model so that its applicability to test dark energy
models is justified as long as such models do not differ much from
a cosmological constant. Nevertheless, this is not the case for
the vector dark energy model whose equation of state varies very
rapidly and, indeed, has a future singularity so that the obtained
3$\sigma$ tension could be due to the dependency of BAO data on
the fiducial model. In any case, this is the less confident
dataset to constraint the vector model and, in general, any dark
energy model, since it may give shifted parameters due to a biased
determination of the sound horizon scale due to the presence of
additional relativistic degrees of freedom, early dark energy or a
non-standard recombination scheme \cite{BAOshifted}.

Finally, CMB data is totally incompatible with flat spatial
sections and, in fact, it predicts a closed universe with a wide
range of $\Omega_m$ allowed. These results show that, contrary to
common belief,  CMB data do not necessarily favors a flat
universe. In Fig. \ref{likelihood}, the corresponding likelihood
for the CMB dataset is plotted and we can see how the flat case is
ruled out for the vector model.

In Fig. \ref{contours} we see that CMB contours are compatible
with BAO at 2$\sigma$ level for small values of $\Omega_m$ and
$\Omega_k$ close to zero. Concerning SNIa contours, CMB is in
conflict with the Union data set contours at more than 3$\sigma$
whereas it is compatible at 1$\sigma$ level with the Gold data
set.

\section{Conclusions}
In this work we have performed a detailed analysis of the
constraints imposed by SNIa, CMB and BAO data on the vector dark
energy model proposed in \cite{cosmicvector}. We have considered
cosmologies with arbitrary spatial curvature and obtained
confidence regions in the $(\Omega_m,\Omega_A)$ plane. We have
found that for the  SNIa Gold dataset, the vector model fit is
better than that of $\Lambda$CDM, but  for the Union dataset the
situation is reversed. We find  that contrary to standard
cosmology, CMB data {\it excludes} a flat universe for this model
and, in fact, predicts a closed spatial geometry. On the other
hand, CMB and SNIa Gold data are perfectly compatible at the
1-sigma level, however SNIa Union dataset exhibits a 3-sigma
tension with CMB. The same level of tension is also found  between
SNIa and BAO measurements, although this may be due to the
dependency of BAO measurements on the fiducial model.

{\em Acknowledgments:}
We would like to thank Eiichiro Komatsu for useful
comments. J.B. is very grateful to the Department of Fisika Teorikoa of
the EHU for their hospitality. This work has been  supported by
DGICYT (Spain) project numbers FPA 2004-02602 and FPA
2005-02327, UCM-Santander PR34/07-15875, CAM/UCM 910309 and
MEC grant BES-2006-12059.

\end{document}